# Electric Field Calculation and PNS Prediction for Head and Body Gradient Coils

Peter B Roemer[1], Trevor Wade[2], Andrew Alejski[2], Koray Ertan[4], Charles A McKenzie[3], Brian K Rutt[4]*

[1]Lutz, FL

[2]Imaging Research Laboratories, Robarts Research Institute, London, ON N6A 5K8, CANADA

[3]Department of Medical Biophysics, Western University, London, ON N6A 5K8, CANADA

[4]Department of Radiology, Stanford University, Stanford, CA 94305, USA

*Correspondence to: Brian K. Rutt, 1201 Welch Road, Stanford, CA 94305, USA. Email: brutt@stanford.edu

## Abstract

**Purpose:** To demonstrate and validate E-field calculation and PNS prediction methods that are accurate, computationally efficient and that could be used to inform regulatory standards.

**Methods:** We describe a simplified method for calculating the spatial distribution of induced E-field over the volume of a body model given a gradient coil vector potential field. The method is easily programmed without finite element or finite difference software, allowing for straightforward and computationally-efficient E-field evaluation. Using these E-field calculations and a range of body models, population-weighted PNS thresholds are determined using established methods and compared against published experimental PNS data for two head gradient coils and one body gradient coil.

**Results:** A head-gradient-appropriate chronaxie value of 669 µs was determined by meta-analysis. Prediction errors between our calculated PNS parameters and the corresponding experimentally measured values were ~5% for the body gradient and ~20% for the symmetric head gradient. Our calculated PNS parameters matched experimental measurements to within experimental uncertainty for 73% of $\Delta G_{min}$ estimates and 80% of $SR_{min}$ estimates. Computation time is seconds for initial E-field maps and milliseconds for E-field updates for different gradient designs, allowing for highly efficient iterative optimization of gradient designs and enabling new dimensions in PNS-optimal gradient design.

**Conclusions:** We have developed accurate and computationally efficient methods for prospectively determining PNS limits, with specific application to head gradient coils.

## INTRODUCTION

Major advances in high resolution, high sensitivity MRI imaging of the brain require further improvements in MR system hardware, especially the gradient system. Present-day gradient systems are operating close to or at the limits of peripheral nerve stimulation (PNS). Whole-body gradients, which are designed for >45cm imaging field-of-view (FOV), demonstrate the lowest PNS thresholds and restrict advanced imaging performance [1]. High-performance head gradient coils have taken on significant recent importance given their intrinsically higher gradient performance and PNS thresholds. Our objective in this work was to develop and validate practical tools for electric-field-based PNS threshold prediction that could be used in a number of ways: 1) to provide input for PNS-optimal gradient design; 2) to understand fundamental aspects of gradient coil PNS; and, 3) to predict PNS threshold parameters for newly developed head gradient coils.

Head gradients can be broadly categorized into two classes, distinguished by the Z-axis symmetry of the magnetic fields they produce (or equivalently by whether or not they produce a $B_0$ concomitant field). The asymmetric class is most common [2-7], while the symmetric class (for example implemented using a folded winding pattern) has also demonstrated promise [8,9]. The PNS properties of a prototype asymmetric head gradient coil (HG2) developed by General Electric [10] have been measured experimentally [11]. Here, we study the electric field (E-field) characteristics of the General Electric ESP research head gradient coil, an asymmetric coil with virtually identical electromagnetic design to HG2 (and because of this equivalence, referred to more generically throughout this paper as the “asymmetric GE” head gradient coil). The Rutt group has designed and built symmetric folded head gradient coils [9] and the PNS thresholds of a prototype symmetric folded head coil known as H3 have been measured [12] making this coil a second logical choice for our study. Despite these and other experimental PNS studies of head gradient coils [1,11-13], there have been no comprehensive analyses of head gradient PNS properties.

While experimental PNS studies are most commonly employed to define PNS limits for a new gradient coil, we believe that PNS threshold prediction via electric field (E-field) calculation is a viable and important alternative to human studies, which are expensive, after the fact, time consuming and not always possible. E-field-based PNS prediction, if performed according to regulatory guidelines, meets safety standards such as IEC 60601-2-33 [14]. More importantly, rapid prediction of PNS thresholds from calculated electric fields could provide the necessary input for PNS-optimal gradient coil design [15], a new and important development in gradient engineering.

In establishing that E-field calculation can serve as an approved method for PNS prediction, the IEC reviewed the physics of pulsed-gradient-induced E-fields in the human body [14]. There is also a substantial body of literature focused on switched-magnetic-coil-induced E-field calculations for PNS threshold prediction [16-27]. It is clear from this background that it is possible to calculate gradient-induced electric fields within the human body by a number of methods, including analytical, finite difference, finite element and specialized methods. These methods all produce accurate E-field estimates, but vary in their computational complexity and speed, and in their abilities to handle body model complexity. It is also clear that the IEC has concluded, at least for body gradient coils, that E-field calculations in simplified uniform-interior body models permit PNS threshold prediction with sufficient accuracy to act as a valid method for establishing PNS safety parameters [14].

In the present work, our aim is to extend the IEC-established E-field-calculation-based method for predicting PNS thresholds, with a specific focus on head gradient coils. We develop computationally-efficient tools to evaluate gradient-induced E-field distributions in realistically sized uniform head-neck-shoulder-torso body models and we predict PNS thresholds for two head gradient coils (one from each of the two major classes) as well as for a reference body gradient coil. The predicted PNS parameters are validated against experimentally-measured results. Our use of simplified models with computationally-efficient PNS threshold prediction opens the door to thorough investigation of PNS properties of head gradient coils over entire populations by simulating E-fields in a distribution of scaled human body models, and allows for generalized PNS-optimal gradient design.

## THEORY

### *Body Model*

We defined realistically sized body models consisting of an ellipsoid to represent the head, connected to neck, shoulders and upper body sections, all with elliptical cross-sections and realistic dimensions. Figure 1 shows the outlines of the body model in the coronal and sagittal planes; the full 3D body model surface is then defined by an ellipse in the transverse (X/Y) plane with semi-major and semi-minor axes (“ellipse radii”) that vary according to these outlines as a function of Z. The Z direction outlines are divided into body / shoulder, neck and head regions. The geometry is described by three radii in the X direction, three radii in the Y direction and five lengths in the Z direction. The body to neck and the neck to head transitions are half periods of a cosine with different lengths, whereas the bottom-of-torso and top-of-head sections are ellipses. The torso section is a straight extension of the shoulder with the same elliptical cross-section; this conservatively large torso section ensures that we capture all currents induced below the shoulders by the head gradient. Table 1 shows the equations that define the body surfaces for each section.

Body model dimensions were obtained from the Humanscale reference manual [28]. The manual provides $2.5^{th}$, $50^{th}$ and $97.5^{th}$ percentile dimensions where the $2.5^{th}$ and $97.5^{th}$ percentiles correspond to +/- 1.96 population standard deviations from the mean and the $50^{th}$ percentile is the median dimension. Table 2 lists the key body / shoulder, neck and head dimensions derived from the Humanscale manual and used to define adult male and female populations of body models according to the equations in Table 1. The total length of the body model from top of head to bottom of body sections (equal to the sum over L0-L4 in Table 1) ranged from 640mm to 658mm. For this work, we generated populations of 100 Gaussian-distributed body models with dimensions sampled from the appropriate distributions for each sex, and these populations were then subjected to E-field calculations. Population-mean PNS threshold parameters (along with their standard deviations) were then computed from these population-wide E-field calculations.

The body /model used for the body gradient coil E-field calculations was an elliptic cylinder with 40cm R/L diameter and 20cm A/P diameter. The 40cm R/L dimension corresponds to the 40cm body model diameter recommendation in the IEC standard, and the 20cm A/P dimension produces a circumference equal to the average shoulder / chest circumference for $50^{th}$ percentile males and females [28].

Our simplified and smooth body models were assumed to have uniform interior electrical properties; compliant with IEC specifications. These simplified body models have a number of advantages, including rapid generation and highly efficient surface E-field calculation. Our central hypothesis was that population-mean PNS thresholds could be accurately estimated using these simplified body models.

*Governing Equations*

The governing equations for E-field in the low frequency magnetoquasistatic regime are well known and described in the IEC safety standard [14,29,30]. The total electric field $\vec{E}$ is given by

$$\vec{E} = -d\vec{A}/dt - \nabla\emptyset \qquad [1]$$

where $\vec{A}$ is the vector potential due to currents and $\emptyset$ is the electrostatic potential due to charge accumulation. The time evolution of charge $\rho$ in a conducting medium with conductivity $\sigma$ and permittivity $\epsilon$ is given by [31]

$$\frac{d\rho}{dt} = \frac{\sigma}{\epsilon}\rho \qquad [2]$$

whose solution is an exponential decay with characteristic time constant $\epsilon/\sigma$. For biological tissues, this time constant is on the order of microseconds [32] and therefore much faster than gradient switching times. As a result, there is no volumetric charge buildup interior to a region of uniform tissue and charges only need to be considered at the interfaces. This means that for a uniform body model, charges only need to be considered on the body surface; this considerably simplifies the analysis. The magnetic fields and reaction electric fields created by the small currents induced in the patient are negligibly small.

The electrical conductivity is zero external to the patient surface and conservation of charge requires a buildup of surface charge to cancel current flowing normal to the surface resulting in a boundary condition at the surface of the body model of

$$\vec{J} \cdot \hat{n} = \sigma\vec{E} \cdot \hat{n} = 0 \qquad [3]$$

where $\hat{n}$ is the unit vector normal to the surface and $\vec{E}$ is the total E-field.

With time-varying gradient magnetic fields, currents induced in the patient and the associated surface charges reach equilibrium in microseconds. The spatial distribution of the E-field is therefore independent of the gradient waveform at the temporal frequencies of interest, while the field amplitude is proportional to the instantaneous time rate of change of the vector potential: $d\vec{A}/dt$. The solution reduces to a static field solution where the input boundary condition is the normal component of $d\vec{A}/dt$ on the surface of the patient model and the resultant total E-field is independent of tissue dielectric constant and conductivity.

*Field Calculations*

E-fields were computed using a compact, targeted and computationally-efficient method, described in detail in the following paragraphs, that is well suited for rapidly evaluating large numbers of body models or gradient designs. Our method generates accurate maps of E-field on the surface of our simplified body models in seconds for initial E-field maps and tens of milliseconds for updated maps where only the gradient coil design is changed.

The vector potential $\vec{A}$ is computed from the known gradient coil winding pattern. To determine the E-field produced by the electrostatic potential Φ, a charge distribution constructed from a set of mathematical basis functions is placed just outside the patient surface. The fields from each basis function of unit amplitude are integrated numerically. The amplitudes of the basis functions are determined by minimizing the E-field normal to the patient surface, as required by Equation [3] in the main paper. The solution converges in the limit as the number of integration points and number of basis functions are increased.

The basis functions for the charge distribution are defined in a surface coordinate system (u,v) where the u coordinate varies from 0 to 360 degrees in the X/Y plane, with u=0 corresponding to the +X axis, and the v coordinate varies from 0 to 1 starting at the inferior end of the body model and ending at the superior end (top of head). Each basis function is constructed as a product of a 1D function in u and a 1D function in v, designated by $U_m(u)\ V_n(v)$. The basis function $U_m(u)$ represents a sine or cosine function depending on the gradient coil axis (X, Y or Z gradient). $V_m(v)$ is piecewise linear in the surface coordinate v (head-foot direction) as shown by the "hat" functions in Figure 2.

The charge density for a single basis function of unit amplitude is of the form

$$q_{m,n}(u,v) = U_m(\mathrm{u})\ V_n(\mathrm{v}) = \alpha_{m,n} \begin{Bmatrix} \sin\left(m\frac{\pi}{180}\ u\right) \\ \cos\left(m\frac{\pi}{180}\ u\right) \end{Bmatrix} V_n(v) \qquad \text{[S1]}$$

Given a surface charge density q, the resultant E-field is determined by integrating Coulomb's law [33] over the surface area in cartesian coordinates

$$\vec{E}(\vec{r}) = \frac{1}{4\pi\varepsilon_o} \int q(\vec{r}^{\,\prime}\ ) \frac{(\vec{r}-\vec{r}^{\,\prime}\ )}{|\vec{r}-\vec{r}^{\,\prime}\ |^{\frac{3}{2}}} dA^{\prime} \qquad \text{[S2]}$$

where $\vec{r}^{\,\prime}$ is the source position at the surface coordinate (u,v). Converting cartesian coordinates to (u,v) surface coordinates provides an expression for the E-field of each basis function

$$\vec{\beta}_{m,n}(\vec{r}) \stackrel{\text{def}}{=} \frac{1}{4\pi\varepsilon_o} \int q_{m,n}(u,v) \frac{(\vec{r}-\vec{r}^{\,\prime}\ )}{|\vec{r}-\vec{r}^{\,\prime}\ |^{\frac{3}{2}}} \left(\frac{ds}{du}\right)\left(\frac{ds}{dv}\right) du\ dv \qquad \text{[S3]}$$

The quantities $\left(\frac{\mathrm{ds}}{\mathrm{du}}\right)$ and $\left(\frac{\mathrm{ds}}{\mathrm{dv}}\right)$ relate a differential distance in the (u,v) surface coordinate system to the global cartesian coordinate system. The total E-field is then the vector sum of the fields from all the basis functions $\sum_{\mathrm{n}} \alpha_{\mathrm{m,n}}\ \vec{\beta}_{\mathrm{m,n}}(\vec{\mathrm{r}})$.

To solve for the unknown amplitudes, $\alpha_{m,n}$ , a set of points is distributed over the patient surface. At each point the normal component of the E-field from the vector potential is added to the static E-field due to charges. The value at each point is squared, weighted by the area associated with that point, and summed over all points. The resultant expression to be minimized is

$$\sum_i W_i^2 \left[\hat{n}_i \cdot \left(\frac{d\vec{A}(\vec{r}_i)}{dt} - \sum_m \sum_n \alpha_{m,n}\, \vec{\beta}_{m,n}(\vec{r}_i)\right)\right]^2 \qquad \text{[S4]}$$

where $W_i$ is the area weight of each point and compensates for nonuniform spacing of the points on the patient surface. The normal to the surface $\hat{n}$, the area weights $W_i$, the vector potential $\vec{A}$ and the field from each basis function $\vec{\beta}$ are determined from the patient geometry and evaluated numerically. The sources are placed on a second surface displaced a small distance by projecting a normal to the patient surface avoiding a division by zero in Equations [S2] and [S3]. Minimization of Equation [S4] involves solving an overdetermined set of linear equations using well-known methods. A measure of convergence is the residual normal component of the electric field on the surface. The exact shape of the second surface and the distance is not critical. This surface simply creates a general electric field over the patient surface where the basis functions amplitudes are numerically determined to meet the boundary condition. As the distance is made larger the required number of required basis function increases in order to meet the boundary condition.

The computation is made more efficient by exploiting gradient field symmetries. For the transverse gradients, six u basis functions were used and limited to odd sine or cosine with m=1,3,5…11. For the Z gradient, the six u basis functions were limited to even sine terms with m=2,4,6…12. For all gradient directions, 50 basis functions were used in the v direction with little benefit from adding more terms. Numerical integrations used 300 points for each u basis function and 16 points for each v basis function. The greater number in u is required because each of the six u basis function spans a full 360 degrees while each of the 50 v basis function only spans $\frac{1}{25}$ of the v direction. The summation in Equation [S4] (over ¼ of the patient surface due to symmetry) comprised 16 points in the u direction and 80 points in the v direction. The source surface was displaced 4 mm outside the surface of the patient model. We used 28,000 filaments to represent each gradient coil conductor path. E-field maps were calculated using 24,000 source points from the basis function and 1,280 points for minimization on ¼ of the body surface. With these sampling parameters, solutions reliably converged with a residual normal E-field component less than 1% of the maximum value of E-field on the surface (see Figure 3), and for this reason we consider the boundary condition defined by Equation [3] in the main paper to have been well met by all solutions.

We assessed the computational efficiency using a laptop with an Intel i7-8550U 1.8GHz 4-core processor, and found that the calculation of full E-field maps required ~17sec and 250MB of RAM per body model and gradient coil. Most commonly, the calculation used only one processor core. Computation time scales with the square of the number of basis functions times the number of summation points in Equation [S4].

Our method differs in important ways from more general finite element or boundary element methods used for MRI gradient coil design and analysis [34-36]. In typical FEM and BEM methods, the surface is sub-divided into a patchwork of elements defined by linear or higher order polynomial segments, through a process of meshing. Our body model is defined very differently, being constructed from piecewise continuous functions in five smooth sections. Although our linear basis functions in the v direction may look like those used in linear FEM or BEM methods, in our case the spatial extent of each basis function is not required to be aligned with any particular meshing of the geometry. Our basis functions in the u direction (angle) extend over the entire circumference which is a single continuous ellipsoidal surface. The integration points are distributed over the body model surface without concern to the start and end points of any element boundaries, but rather according to the mathematical model of the surface. Our method provides a compact, targeted and highly efficient solution to computing gradient-induced E-fields. Our method can be easily integrated into gradient design codes, which are often similarly customized to the gradient design problem.

*Calculation of PNS Thresholds from E-fields*

The IEC 60601-2-33 safety standard [14] allows for two different methods for PNS threshold determination for head gradient coils: 1) E-field calculations with use of IEC-specified factors to compute PNS threshold parameters from peak E-fields on the surface of a uniform-interior body model; and, 2) experimental PNS measurements in human subjects. A third method using dB/dt calculation is currently permitted by the IEC for use with body gradients, but not head gradients.

The E-field calculation method is based on the modeling of the PNS threshold according to a strength-duration relationship defining the minimum E-field applied for duration Δt. PNS model factors for this strength-duration relationship are rheobase rb (minimum E-field to cause nerve stimulation) and chronaxie ch (time constant for nerve stimulation). IEC requires that for body gradient coils, fixed values of rb = 2.2 V/m and ch = 360 μs be used for calculating PNS thresholds from E-fields. Conversion of the E-field strength-duration relationship to the linear magnetostimulation formulation [13], where PNS parameters are given by the intercept and slope, yields

$$\Delta G_{stim} = \Delta G_{min} + \Delta t \, SR_{min} \quad [4]$$

The PNS parameters $\Delta G_{min}$ and $SR_{min}$ are the minimum gradient excursion to cause stimulation at any switching time and the minimum slew rate to cause stimulation at any gradient amplitude, respectively, and are given by [13]

$$\Delta G_{min} = \frac{\mathrm{rb}}{\mathrm{E_{max}}/SR}\,\mathrm{ch}$$
$$SR_{min} = \frac{\mathrm{rb}}{\mathrm{E_{max}}/SR} \quad [5]$$

where $E_{max}/SR$ is the maximum value of electric field per unit slew rate found on the body surface.

In the present work, Equations [5] are used to calculate PNS parameters $\Delta G_{min}$ and $SR_{min}$ following calculation of E-fields for a given gradient coil and individual body model. We compute population-weighted mean PNS parameters by averaging over 100 normally-distributed body models for each sex, which are then averaged together (with equal weighting reflecting near-equal male and female populations) to yield the overall population mean PNS parameters.

In addition to using the IEC-specified chronaxie for body gradient coils, we explored one variation of the IEC method for estimating PNS threshold parameters by E-field calculation: use of a longer value of chronaxie for head gradient coils. Longer chronaxie values have consistently been measured for head gradient coils compared to body gradient coils [1,11-13]. We derived a single candidate chronaxie value by averaging seven chronaxie values extracted from the known literature containing head gradient PNS measurements. These seven measurements correspond to three different head gradient coils: one coil (GE HG1) with one gradient direction measured [11], one coil (GE MAGNUS) with two gradient directions measured [37], and one coil (Siemens AC84) with four gradient directions measured [38]. The seven chronaxie values were 458, 766, 620, 781, 533, 879, 648 µs, respectively, with an average value of 669 µs and a standard deviation of 148 µs. It should be noted that the chronaxie values for the Siemens AC84 were derived by extracting $\Delta G_{min}$ (intercept) and $SR_{min}$ (slope) from the measured PNS thresholds shown in Davids et al [38]; chronaxie was then calculated as $\Delta G_{min}$ / $SR_{min}$ according to Equation [5]. It should also be noted that these seven measurements were made using different gradient coils than those used for validation of the E-field-based PNS predictions in the present work, which means that this meta-analysis produced a completely independent estimate of head-gradient-appropriate chronaxie. We then predicted

PNS threshold parameters in two ways: 1) using IEC-established body gradient factors; and, 2) using our head-gradient-appropriate chronaxie (669 μs) for head gradient coils, with otherwise unmodified IEC factors and methods.

## METHODS

We examined one coil from each of the two major classes of head gradient coil (asymmetric and symmetric); this provided the widest possible range of PNS behavior for our validations. These two head gradient coils were also the only two coils for which both winding patterns and experimental PNS measurements were available to us.

All E-field calculations for the asymmetric head gradient class utilized an asymmetric head gradient design (GE ESP) with the same 26 cm diameter imaging volume and nearly identical EM and physical properties to the GE HG2 head gradient described in the literature [10]; the PNS properties of these two coils should be virtually identical. For the experimental PNS study of HG2, the gradient coil was driven to a maximum gradient strength $G_{max}$ of 85 mT/m and maximum slew rate $S_{max}$ of 700 T/m/s [11]. PNS measurements were attempted for the three principal gradient directions (X, Y, Z), with data reported for one of these (X); we show the reported PNS parameters $\Delta G_{min}$ and $SR_{min}$ for the HG2 X-axis in Table 3, along with the standard errors of these mean estimates. It should be noted that the reported coefficients of variation (standard error of the mean divided by mean) for these HG2 PNS threshold parameter estimates were relatively high: 32% for $\Delta G_{min}$ and 66% for $SR_{min}$. According to the original article [11], these relatively large errors are due to the fact that the hardware limits prevented experimental data sampling outside a relatively small portion of the stimulation curve.

The symmetric folded head gradient class has been described in the literature [9,12,39]. The H3 symmetric folded head gradient was designed to be rapidly insertable into whole-body GE clinical scanners and was designed with 22 cm (transverse) by 19 cm (Z) imaging volume diameters. For this H3 gradient design, experimental PNS measurements were made in 15 adult human subjects (6 female, mean age 33.3 yrs, mean weight 78.7 kg), using the following hardware limits: $G_{max}$ 120 mT/m and $S_{max}$ 1200 T/m/s [12]. This experimental PNS study acquired data using all combinations of principal directions (X, Y, Z, XY, XZ, YZ, XYZ), as well as two subject positions: shoulder-coil contact and 2 cm shoulder-coil gap. Because the original publication presented only some of the key PNS results, we provide the complete set of measured PNS parameter estimates along with their standard errors here, in Tables 4 and 5.

We also performed E-field calculations and PNS predictions for a “reference” whole body gradient coil – the GE BRM gradient coil – to act as further validation of our methods. This is a conventional large-field-of-view body gradient coil whose design was available to us and whose PNS threshold parameters have been measured by our group with consistent experimental methods and analysis [1]. This experimental PNS study reported data for two directions (Y, XY) with each subject positioned for lowest PNS threshold; we show these previously reported PNS parameter estimates along with their standard errors in Table 6.

PNS threshold parameters depend on gradient design, gradient direction and body position. For purposes of direct comparison to experimental studies, which report population means, we used our E-field calculations together with IEC-defined rheobase and chronaxie [14] to estimate the equivalent population-weighted PNS parameters (along with the standard errors of these estimates) for all gradient coils, gradient directions and body positions for which experimental data were available. We computed percent differences between calculated and experimental data for both $\Delta G_{min}$ and $SR_{min}$. Finally, we assessed the accuracy of our predicted PNS parameters according to mean absolute error (MAE). We characterized the overall accuracy of our E-field-based PNS prediction by averaging MAE across all reported measurements.

In select cases, we validated our E-field calculations by numerical finite element modeling using an established EM modeling package (Sim4Life, ZMT, Zurich) with the same coil and model geometries.

## RESULTS

Figure 4 shows the measured PNS data (including individual subject data points) for the H3 gradient coil for all gradient directions and both subject positions. Intercept $\Delta G_{min}$ and slope $SR_{min}$ were extracted by linear regression fitting of the mean $\Delta G_{stim}$ versus risetime $\Delta t$ data (with logistic regression used to estimate this mean $\Delta G_{stim}$ at each $\Delta t$ according to the methods described by Zhang et al. [1]); the resulting estimates of PNS parameters $\Delta G_{min}$ and $SR_{min}$ are tabulated in Tables 4 and 5 (corresponding to the two body positions) along with standard errors of these estimates.

Figure 5 shows the magnitude of the normalized E-field (units of mV/m per T/m/s) on the surface of the 50$^{th}$ percentile male body model for the asymmetric GE and symmetric H3 head gradient coils and on a 40cm x 20 cm elliptic cylinder for the BRM body gradient coil, showing all seven gradient directions for each coil. The number shown in each panel indicates the maximum value found on the surface. The vectors on the surface show the direction of the electric fields / currents.

Figures 5a and 5b show that the transverse windings of the asymmetric GE versus symmetric H3 head gradient coils produce E-field hotspots with very different spatial distributions: for the asymmetric GE coil, peak E-fields for X and Y windings are located at the top of the head whereas for the symmetric H3 coil, X and Y windings produce peak E-fields in the neck / shoulders region. The E-field distributions for the Z windings are more similar between the two coil classes; this is expected because both coils use symmetric Z-gradient designs. However, the amplitude of the symmetric H3 Z coil's E-field is considerably lower than that of the asymmetric GE coil, a result of the smaller linearity region of the symmetric H3 coil.

From Figure 5a, we note that the locations of maximum E-field for the asymmetric GE head gradient, being the face and top of the head for the X gradient and the shoulder and back areas for the Z gradient, are in approximate agreement with the locations for the onset of sensation reported in the HG2 experimental study [11], although precise correlations were not possible given the limited experimental data relating to stimulation location. Correspondingly, the locations of maximum E-field for the symmetric H3 head gradient, being lateralized to the sides of the neck and shoulders for the Y gradient, are in approximate agreement with the experimentally reported locations from the H3 study (~80% of Y gradient PNS sensations were lateralized to the R/L sides of the head, neck, shoulders) [12].

From Figure 5c, it is seen that for the BRM body gradient the maximum surface E-fields are located on the right / left sides of the elliptic cylindrical volume, and ~40 cm inferior to isocenter. E-field amplitudes are highest for the Y gradient direction, with nearly double the peak value compared to the X or Z directions; this predicts that PNS will occur first with the Y gradient direction. This result is indeed confirmed by experimental PNS studies [1]. Interestingly, the XY oblique gradient direction shows the second-highest value of peak E-field, with hotspot located near one side but rotated somewhat away from the right/left axis, while still located ~40 cm inferior to isocenter. This result agrees with experimental PNS measurements showing that the BRM XY coil typically stimulates off-center (R/L) in the buttocks / lower back region of adult human subjects [1,40].

Tables 3-6 show experimentally measured PNS threshold parameters $\Delta G_{min}$ and $SR_{min}$ (mean and standard error), along with the corresponding calculated values (mean and standard deviation), and percent differences between calculated and experimental mean values, for all gradient directions. Table 3 shows results for the asymmetric GE head gradient, while Tables 4 and 5 show results for the symmetric H3 head gradient in two body positions: shoulder-coil contact (brain centered) and 2cm shoulder-coil gap, respectively. Table 6 contains results for the BRM body gradient coil.

For the head gradient coils, surface E-field distributions were calculated as described: $E_{max}$/SR (peak magnitude of E field found on the surface normalized by slew rate) was extracted for each of 100 normally-distributed body models, and population-weighted PNS parameters were estimated using Equation [5] and IEC-specified scale factors (rheobase rb of 2.2 V/m and chronaxie ch of 360 µs). Columns were added to the head gradient coil calculations (Tables 3-5) to show $\Delta G_{min}$ recalculated using the head-gradient-appropriate chronaxie value of 669 µs. For the body gradient coil, calculations used the single elliptic cylindrical body model, but otherwise followed the same procedure to estimate PNS parameters.

The results in Tables 3-6 indicate correct transverse gradient PNS threshold orderings, in agreement with the known PNS characteristics of the corresponding gradient coils: the asymmetric GE head gradient design is known to have lower X compared to Y PNS thresholds, in agreement with experiment [11], whereas the opposite is seen for H3 and BRM, again in agreement with experiment [1,12].

Tables 3-6 also show percent errors between calculated and experimentally measured PNS parameters. For the reference body gradient coil (BRM), calculated PNS threshold parameters match the experimentally measured equivalents for the X and XY gradient directions (the only two reported gradient directions [1]) to within 10% MAE. Averaging across these two gradient directions, the discrepancies between computed and measured values are 4.6% MAE for $\Delta G_{min}$ and 4.4% MAE for $SR_{min}$. This strong agreement represents a first validation of our computational methods.

Examining the symmetric H3 head gradient results in Tables 4 and 5, using the chronaxie value of 360 µs and considering all measured gradient directions at the two body positions, our calculated PNS threshold parameters differed on average from their experimentally measured equivalents by 40% MAE for $\Delta G_{min}$ and 23% MAE for $SR_{min}$. The 40% discrepancy for $\Delta G_{min}$ was reduced to 14% MAE if we used the head-gradient-appropriate chronaxie value of 669 µs instead of 360 µs. Overall, the agreement between calculated and measured PNS parameters is reasonable (20% MAE) assuming the longer chronaxie value of 669 µs is used in the conversion of E-fields to PNS parameters. This qualitative and quantitative match between calculated and experimental PNS characteristics for the symmetric folded head gradient class represents a second strong validation of our computational methods.

For the asymmetric GE comparison results shown in Table 3 and using the chronaxie value of 360 µs, calculated PNS threshold parameters differ from experimentally measured values by 7.3% for $\Delta G_{min}$ and 58% for $SR_{min}$. If the longer chronaxie value of 669 µs is used instead of 360 µs, the error in calculated $\Delta G_{min}$ *increases* to 72% while the error in calculated $SR_{min}$ doesn't change (the latter being independent of chronaxie). In other words, the prediction of PNS parameters for the asymmetric GE coil X axis is not very accurate and gets substantially worse with the proposed modified chronaxie.

## DISCUSSION

PNS strongly limits present-day MRI systems. Gradient coils with smaller imaging regions lead to reduced magnetic flux intercepted by the patient, yielding higher PNS thresholds. Therefore head gradient coils, with their smaller imaging regions, have taken on new importance in the drive toward higher performance MR imaging of the brain. However, there have only been a few experimental studies of the PNS properties of head gradient coils in the literature [1,11,12,37,38] and no fundamental / theoretical studies as to the best head gradient coil design to further reduce E-fields for the full range of patient sizes. The computational methods presented here provide the basis for such fundamental analyses. Our calculations have already provided new insights into the E-field distributions for different head gradient coil designs.

Overall, our calculations produced reasonably accurate predictions of PNS thresholds as demonstrated by direct validation against three different experimental PNS studies comprising a total of 30 distinct PNS measurements (counting both PNS parameters from all coils, all gradient directions and all body positions that yielded measurable results). For the BRM body gradient, the PNS parameter prediction errors were very low – less than 5%. For the symmetric H3 head gradient, the overall PNS parameter prediction errors were reasonably low – in the range of 32% using 360µs chronaxie but less than 20% using 669 µs chronaxie. For the asymmetric GE head gradient coil, the PNS parameter prediction errors were considerably higher – in the range of 60-70%.

We compared the discrepancies between calculated and experimental PNS parameters to the experimental uncertainty in those parameters (defined as +/-3CV where CV is the coefficient of variation of these estimated parameters) and found that when using 360 µs chronaxie for all coils, these discrepancies were within experimental uncertainty for only 20% (3 out of 15) of the $\Delta G_{min}$ estimates. On the other hand, when using the head-gradient-appropriate chronaxie value of 669 µs, these discrepancies were within experimental uncertainty for 73% (11 out of 15) of the experiments. For $SR_{min}$ estimates (which are independent of chronaxie), discrepancies were within experimental uncertainty for 80% (12 out of 15) of the experiments. We conclude that our E-field-based calculations enable the prediction of experimentally measured PNS parameters to within experimental error for a majority of experiments, assuming use of a head-gradient-appropriate chronaxie value.

For the asymmetric GE coil, we believe that the large discrepancies between calculated and measured PNS threshold parameters are due to the limited range over which PNS measurements could be made with the HG2 gradient within the confines of its hardware limits, which led to the large experimental uncertainties in $\Delta G_{min}$ (32% CV) and $SR_{min}$ (66% CV) [10]. In future, higher current gradient drivers should allow more precise measurement of PNS parameters for the X gradient direction of the asymmetric GE head gradient coil and should permit measurement of PNS thresholds for other gradient directions.

We also note that for the asymmetric GE calculations, the Z axis PNS threshold is predicted to be similar if not slightly lower than that of the X axis; on the other hand, the experimental study found some Z axis stimulation but less than what was found for the X axis. Our calculated PNS thresholds for both X and Z axes of the asymmetric GE head gradient were close to or above existing hardware limits, and for this reason we suspect that the experimental measurements are not reliable enough at present to demonstrate the relative ordering of X and Z thresholds. Secondly, based on our analysis of E-fields in this situation, we believe that PNS thresholds for the Z axis are strongly dependent on body dimensions in the neck / shoulder region, which have greater population standard deviation when compared to head dimensions. There could also be differences between our body model distribution and the actual subject population used for the HG2 experimental PNS study, which would constitute another factor contributing to the observed discrepancies.

For the H3 gradient coil, the largest discrepancies between calculated and experimental PNS parameters were seen for the slope parameter $SR_{min}$, with discrepancies up to 80% found for one gradient direction (H3 X). For all other measurements, discrepancies were much lower: 30% or lower when the head-gradient-appropriate chronaxie value was used. We believe that measurements of PNS parameters for the H3 X axis suffer from the problem of error inflation that is known to occur when the PNS threshold barely intersects the hardware-limited region. In such a case, the number of points that contribute to the linear fit of the $\Delta G_{stim}$ vs $\Delta t$ data drops to a small number of closely spaced $\Delta t$ values – this causes the experimental uncertainty in $\Delta G_{min}$ and $SR_{min}$ (intercept and slope) to increase, sometimes drastically [1]. Examining Figure 4, we see that considering all measurable gradient directions the H3 X and to a lesser extent the H3 XZ PNS thresholds come closest to just skimming the corner of the hardware-limited region; this behavior is similar to that of the GE HG2 X [10]. Furthermore, examining the standard errors of the experimentally measured PNS parameter estimates in Tables 3, 4 and 5, we find larger coefficients of variation for HG2 X, H3 X and H3 XZ compared to other measurements. Given that HG2 X, H3 X and H3 XZ correspond to the gradient coils/directions that demonstrate most of the large discrepancies between calculated and measured PNS thresholds, these findings support the conclusion that these largest discrepancies result from error inflation caused by PNS thresholds barely intersecting the hardware-limited region. With the exception of these special cases, our calculated PNS parameters predominately match measured PNS parameters to within experimental uncertainty and with reasonable accuracy over a range of gradient coils and directions.

Overall, we find that our E-field calculations combined with the IEC-specified values of both rheobase and chronaxie predict PNS threshold parameters with high accuracy for the body gradient and reasonable accuracy for the symmetric H3 head gradient. We also show that by modifying chronaxie to the head-gradient-appropriate value of 669 µs without changing the IEC-mandated value for rheobase (or in fact changing any other IEC specifications), we achieve significantly higher prediction accuracy for the H3 head gradient. Therefore, a major result of our work is that it is not necessary to diverge very far from the IEC-mandated values to achieve reasonably accurate PNS prediction for head gradient coils using the standardized E-field calculation strategy we have laid out.

As mentioned above, to maximize prediction accuracy on the estimation of $\Delta G_{min}$, we found it necessary to use a longer chronaxie value for head gradient coils compared to the IEC-specified value of 360 µs used for body gradient coils. Multiple PNS studies have measured chronaxie values for head or small-imaging-region body gradient coils that are significantly longer than 360 µs [1,11-13,37,38] so it not a surprise that accurate prediction of $\Delta G_{min}$ from E-fields for head gradient coils would require the use of longer chronaxie values. We do not address here the question of *why* chronaxie values are different for body and head gradient coils. This remains an interesting but unresolved question that has been commented on and investigated in prior literature [1,40-42]; but the fact that measured chronaxie values are different between body and head gradients is undeniable, and our results provide further evidence in support of this. Further work to elucidate this dependence of measured chronaxie on imaging region size or other factors is warranted.

In our evaluation of the BRM whole-body gradient, we determined that the IEC-specified 360 µs chronaxie value was appropriate. However, we needed to use a more realistic 40cm x 20cm ellipsoidal cross-section body model to achieve correct PNS predictions for multiple gradient directions. We note that the circumference and aspect ratio of this elliptical cross-section body model closely approximates that of the average Humanscale body in the shoulder / chest region [28].

Based on these results, we would recommend that the appropriate regulatory safety standards committee consider allowing the use of longer chronaxie values in conjunction with E-field calculations to predict PNS threshold parameters for specialty head gradient coils. We are suggesting here that the IEC-specified body-gradient-appropriate value of 360 μs be replaced by a single longer head-gradient-appropriate value for use when E-field-based PNS threshold prediction is required for head gradients.

The E-field theory and body models described here conform to regulatory requirements; this combined with the reasonable prediction accuracy that we have demonstrated here means that these methods could substitute for experimental PNS studies, which are expensive, slow and often limited to small numbers of human subjects. Furthermore, experimental PNS studies may not always be possible, for example in vulnerable populations such as children. An accurate computational approach would represent an important alternative for obtaining regulatory and ethical approval for the use of novel gradient technologies in those vulnerable populations.

A graphical display of PNS limits overlaid onto hardware limits for all H3 gradient directions and body positions, comparing calculations to experimental measurements and using 669 μs chronaxie for the head gradient calculations, is shown in Figure 4. This figure shows the reasonable match between predicted and measured PNS thresholds; it also demonstrates the challenge of measuring some head gradient coil PNS limits, especially those that lie close to the hardware limits. This reinforces the value of making PNS measurements with combined directions such as XY, XZ, YZ, XYZ. We used these combined directions for our H3 measurements and showed that high precision PNS parameter estimation is often possible only by using these combined gradient directions as a result of their larger achievable field amplitudes. Having said this, the H3 XZ direction is an example of a combined gradient direction that still barely intersects the hardware-limited region.

We would also point out that there may be other factors apart from experimental uncertainty that explain the remaining discrepancies between our calculated-E-field-based predictions and experimental measurements of PNS parameters. One has been mentioned already: a sampling bias that would occur if the anatomical dimensions of the small cohort of subjects used in any given experimental study are not consistent with a random sample from the Gaussian-distributed population that our calculations used. Another factor is that our simplified body models cannot possibly represent the full anatomical and physiological complexity of the real human body; this factor has motivated a number of groups to use more detailed body models for switched-magnetic-field-induced E-field calculations [15,25,38,43-49]. Despite this, our finding that PNS threshold predictions can be made with an average absolute error of

~5% for a typical body gradient coil and ~20% for a typical head gradient coil means that our simplified body models have important application in PNS and head gradient research.

While prior literature has described E-field calculations to predict PNS thresholds and stimulation locations for switched gradient coils [16-21], the combination of computational speed, direct validation with reasonable accuracy compared to matched experimental measurements from head and body gradient coils, and the regulatory compatibility of our methods, represents an advance over that prior work. Our work demonstrates an important capability for predicting population-mean PNS thresholds with sufficient accuracy to be useful for a number of important applications in head gradient design and development.

Recent work by Davids and others [25,38,50-52] has used anatomically- ***and*** compositionally-realistic body models together with physiologically-realistic neuronal activation models to predict PNS thresholds for individual body models. The use of realistic body and neuronal activation models is clearly important for understanding PNS fundamentals and for prediction of both threshold *and location* of peripheral nerve stimulation at the level of the individual human subject. As a concrete example, we believe that such advanced modeling may be necessary to explain the differences in measured chronaxie between head and body gradient coils. Compared to these methods, our work is more narrowly focused on building alternative tools for prediction of *population-mean* PNS thresholds that are compatible with present-day regulatory guidelines for establishing PNS safety of gradient coils. The uniform-interior body models that we used are compatible with established MRI safety standards such as IEC 60601-2-33 [14,29,30]. Use of non-uniform body models modulates the E-fields yielding local E-field intensifications that can complicate the interpretation and lead to lower computed PNS thresholds compared to the population mean values obtained with an experimental human study.

Finally, the high computational efficiency of our methods, which produce E-field maps over the entire body model in seconds with updates for different gradient designs in *tens of milliseconds*, allows for practical PNS-constrained optimization of gradient designs. While the concept of PNS-constrained gradient design is not new [36,53,54], it is one that has been advanced recently by Davids et al [15] and ourselves [55]. We believe that the fast and accurate E-field calculations and PNS prediction tools described here form the ideal engine to drive such PNS-optimal gradient design.

## CONCLUSION

We have developed new computational methods for the evaluation of switched-gradient-induced electric fields in simplified body models. We have used these calculations to predict PNS parameters for three different gradient coils, including two classes of head gradient coils. Our predictions proved to be reasonably accurate as assessed by direct comparison to experimental PNS measurements using the same three gradient coils. The speed and accuracy of the newly developed methods motivate us to integrate E-field calculations into existing design software, allowing the inclusion of E-field constraints as part of gradient coil design optimization.

## Acknowledgements

We wish to thank GE Healthcare and Jonathan Murray for the opportunity to conduct the research required for publication of this article, and for the ESP and BRM gradient coil information used in this work. We gratefully acknowledge the contributions of Dr. Graeme McKinnon to the symmetric folded H3 head gradient design effort. We acknowledge the support of ZMT (Zurich MedTech) who provided expertise and access to Sim4Life for electromagnetic field modeling.

Grant support: NIH P41 EB015891, NIH R01 EB025131, NIH U01 EB025144

# TABLES

**Table 1**

Body model surface shape equations.

| Section | Equation of Surface |
|---|---|
| Body End | $\frac{x^2}{R_{x0}^2}+\frac{y^2}{R_{y0}^2}+\frac{z^2}{L_0^2}=1$ |
| Body | $\frac{x^2}{R_{x0}^2}+\frac{y^2}{R_{y0}^2}=1$ |
| Body/Neck Transition | $z_2 \equiv Z_{BC}-L_2-L_3-L_4$ <br> $R_x \equiv \frac{(R_{x1}+R_{x0})}{2}-\frac{(R_{x1}-R_{x0})}{2}cos\left(\frac{z-z_2}{L_2}\pi\right)$ <br> $R_y \equiv \frac{(R_{y1}+R_{y0})}{2}-\frac{(R_{y1}-R_{y0})}{2}cos\left(\frac{z-z_2}{L_2}\pi\right)$ <br> $\frac{x^2}{R_x^2}+\frac{y^2}{R_y^2}=1$ |
| Neck/Head Transition | $z_3 \equiv Z_{BC}-L_3-L_4$ <br> $R_x \equiv \frac{(R_{x2}+R_{x1})}{2}-\frac{(R_{x2}-R_{x1})}{2}cos\left(\frac{z-z_3}{L_3}\pi\right)$ <br> $R_y \equiv \frac{(R_{y2}+R_{y1})}{2}-\frac{(R_{y2}-R_{y1})}{2}cos\left(\frac{z-z_3}{L_3}\pi\right)$ <br> $\frac{x^2}{R_x^2}+\frac{y^2}{R_y^2}=1$ |
| Top of Head | $\frac{x^2}{R_{x2}^2}+\frac{y^2}{R_{y2}^2}+\frac{z^2}{L_4^2}=1$ |

**Table 2**

Body model values used in E-field calculations (mm units). Displacements of brain center and eye center from top of head are included (bottom two rows), showing that brain center is located slightly superior to eye center as expected. The term "ellipse radii" refers to semi-major and semi-minor axis dimensions.

| Item | Symbol | Male | | | Female | | |
|---|---|---|---|---|---|---|---|
| | | **2.5%** | **50%** | **97.5%** | **2.5%** | **50%** | **97.5%** |
| x ellipse radii shoulder | $R_{x0}$ | 203.0 | 225.0 | 246.5 | 183.0 | 203.0 | 225.0 |
| x ellipse radii neck | $R_{x1}$ | 53.5 | 58.5 | 63.5 | 51.7 | 55.9 | 60.2 |
| x ellipse radii head | $R_{x2}$ | 72.5 | 77.5 | 82.5 | 72.5 | 72.5 | 77.5 |
| y ellipse radii torso | $R_{y0}$ | 98.0 | 114.5 | 136.0 | 97.0 | 107.0 | 117.0 |
| y ellipse radii neck | $R_{y1}$ | 57.2 | 61.2 | 66.7 | 51.7 | 55.9 | 60.2 |
| y ellipse radii head | $R_{y2}$ | 92.5 | 98.0 | 104.0 | 91.5 | 92.5 | 98.0 |
| z lengths (ellipse radii) body endcap | $L_0$ | 50.0 | 50.0 | 50.0 | 50.0 | 50.0 | 50.0 |
| z lengths body straight section | $L_1$ | 252.3 | 235.0 | 219.0 | 267.3 | 249.0 | 231.8 |
| z lengths body/neck transition | $L_2$ | 150.5 | 164.0 | 173.0 | 142.5 | 157.0 | 170.5 |
| z lengths neck/head transition | $L_3$ | 97.3 | 101.0 | 108.0 | 90.3 | 94.0 | 97.8 |
| z lengths (ellipse radii) top of head | $L_4$ | 97.3 | 101.0 | 108.0 | 90.3 | 94.0 | 97.8 |
| z brain center (relative to top of head) | $Z_{BC}$ | -97.3 | -101.0 | -108.0 | -90.3 | -94.0 | -97.8 |
| z eye center (relative to top of head) | $Z_{EC}$ | -102.0 | -112.0 | -119.0 | -102.0 | -112.0 | -119.0 |

**Table 3**
Experimentally measured mean and standard error, calculated mean and population standard deviation, and percent difference between means, for PNS parameters $\Delta G_{min}$ (mT/m units) and $SR_{min}$ (T/m/s units) for the asymmetric GE gradient coil. $\Delta G_{min}$ calculated with both 360µs and 669µs chronaxie values.

| | Experimental | | | | Calculated | | | | | | Percent Difference | | |
|---|---|---|---|---|---|---|---|---|---|---|---|---|---|
| | ΔGmin T/m/s | | SRmin T/m/s | | (360µs) ΔGmin mT/m | | (669µs) ΔGmin mT/m | | SRmin T/m/s | | (360µs) ΔGmin mT/m | (669µs) ΔGmin mT/m | SRmin T/m/s |
| Gradient Direction | mean | se | mean | se | mean | sd | mean | sd | mean | sd | mean | mean | mean |
| X | 99 | 31 | 161 | 106 | 91 | 5.6 | 152 | 9.4 | 254 | 15.7 | -7.3% | 54.6% | 57.8% |
| Y | | | | | 111 | 6.8 | 185 | 11.4 | 308 | 19.0 | | | |
| Z | | | | | 86 | 3.9 | 144 | 6.5 | 240 | 10.8 | | | |
| XY | | | | | 101 | 6.2 | 169 | 10.4 | 281 | 17.4 | | | |
| XZ | | | | | 114 | 7.9 | 190 | 13.1 | 317 | 21.9 | | | |
| YZ | | | | | 106 | 4.8 | 177 | 8.0 | 295 | 13.3 | | | |
| XYZ | | | | | 119 | 7.6 | 198 | 12.6 | 329 | 21.0 | | | |

**Table 4**
Experimentally measured mean and standard error, calculated mean and population standard deviation, and percent difference between means, for PNS parameters $\Delta G_{min}$ (mT/m units) and $SR_{min}$ (T/m/s units) for the symmetric H3 gradient coil with shoulder-coil contact (brain centered). $\Delta G_{min}$ calculated with both 360µs and 669µs chronaxie values.

| | Experimental | | | | Calculated | | | | | | Percent Difference | | |
|---|---|---|---|---|---|---|---|---|---|---|---|---|---|
| | ΔGmin mT/m | | SRmin T/m/s | | (360µs) ΔGmin mT/m | | (669µs) ΔGmin mT/m | | SRmin T/m/s | | (360µs) ΔGmin mT/m | (669µs) ΔGmin mT/m | SRmin T/m/s |
| Gradient Direction | mean | se | mean | se | mean | sd | mean | sd | mean | sd | mean | mean | mean |
| **X** | 201 | 29.9 | 152 | 92.6 | 98 | 9.9 | 183 | 18.5 | 273 | 27.6 | -51.1% | -9.1% | 79.7% |
| **Y** | 101 | 3.8 | 183 | 10.3 | 71 | 4.6 | 132 | 8.5 | 197 | 12.7 | -29.6% | 30.9% | 7.7% |
| **Z** | | | | | 139 | 4.0 | 259 | 7.4 | 387 | 11.0 | | | |
| **XY** | 129 | 4.3 | 177 | 6.0 | 80 | 6.4 | 150 | 11.9 | 223 | 17.9 | -37.6% | 16% | 26.3% |
| **XZ** | 211 | 25.6 | 341 | 97.1 | 120 | 9.2 | 223 | 17.1 | 333 | 25.6 | -43.1% | 5.7% | -2.3% |
| **YZ** | 150 | 8.2 | 331 | 28.5 | 94 | 5.4 | 174 | 10.0 | 261 | 14.9 | -37.6% | 16.1% | -21.2% |
| **XYZ** | 175 | 5.8 | 228 | 9.7 | 89 | 6.0 | 165 | 11.1 | 246 | 16.5 | -49.3% | -5.8% | 8.3% |

**Table 5**
Experimentally measured mean and standard error, calculated mean and population standard deviation, and percent difference between means, for PNS parameters $\Delta G_{min}$ (mT/m units) and $SR_{min}$ (T/m/s units) for the symmetric H3 gradient coil with 2cm shoulder-coil gap. $\Delta G_{min}$ calculated with both 360μs and 669μs chronaxie values.

| | Experimental | | | | Calculated | | | | | | Percent Difference | | |
|---|---|---|---|---|---|---|---|---|---|---|---|---|---|
| | ΔGmin mT/m | | SRmin T/m/s | | (360μs) ΔGmin mT/m | | (669μs) ΔGmin mT/m | | SRmin T/m/s | | (360μs) ΔGmin mT/m | (669μs) ΔGmin mT/m | SRmin T/m/s |
| **Gradient Direction** | **mean** | **se** | **mean** | **se** | **mean** | **sd** | **mean** | **sd** | **mean** | **sd** | **mean** | **mean** | **mean** |
| **X** | 160 | 10.9 | 154 | 26.6 | 99 | 9.2 | 183 | 17.1 | 274 | 25.5 | -38.5% | 14.3% | 78.1% |
| **Y** | 106 | 5.9 | 230 | 24.9 | 73 | 4.1 | 136 | 7.6 | 203 | 11.3 | -31.0% | 28.2% | -11.7% |
| **Z** | | | | | 163 | 3.3 | 304 | 6.2 | 454 | 9.2 | | | |
| **XY** | 141 | 7.4 | 222 | 18.0 | 82 | 5.7 | 152 | 10.6 | 227 | 15.8 | -41.9% | 7.9% | 2.4% |
| **XZ** | 179 | 9.6 | 275 | 32.6 | 126 | 10.3 | 233 | 19.1 | 349 | 28.6 | -29.7% | 30.6% | 26.7% |
| **YZ** | 194 | 10.6 | 280 | 27.2 | 98 | 5.7 | 182 | 10.5 | 272 | 15.8 | -49.5% | -6.2% | -2.9% |
| **XYZ** | 175 | 9.6 | 289 | 26.7 | 92 | 6.2 | 171 | 11.4 | 256 | 17.1 | -47.3% | -2.1% | -11.3% |

**Table 6**

Experimentally measured mean and standard error, calculated mean and population standard deviation, and percent difference between means, for PNS parameters $\Delta G_{min}$ (mT/m units) and $SR_{min}$ (T/m/s units) for the BRM body gradient coil. $\Delta G_{min}$ was calculated with 360µs chronaxie value.

| | **Experimental** | | | | **Calculated** | | **Percent Difference** | |
|---|---|---|---|---|---|---|---|---|
| | **ΔGmin mT/m** | | **SRmin T/m/s** | | **ΔGmin mT/m** | **SRmin T/m/s** | **ΔGmin mT/m** | **SRmin T/m/s** |
| **Gradient Direction** | **mean** | **se** | **mean** | **se** | **mean** | **mean** | **mean** | **mean** |
| **X** | | | | | 35.8 | 99 | | |
| **Y** | 20.9 | 4.2 | 59.5 | 7.5 | 19.7 | 55 | -5.6% | -8.0% |
| **Z** | | | | | 33.8 | 94 | | |
| **XY** | 24.7 | 1.9 | 66.8 | 3.5 | 23.8 | 66 | -3.6% | -0.8% |
| **XZ** | | | | | 34.8 | 97 | | |
| **YZ** | | | | | 26.4 | 73 | | |
| **XYZ** | | | | | 26.9 | 75 | | |

## FIGURES

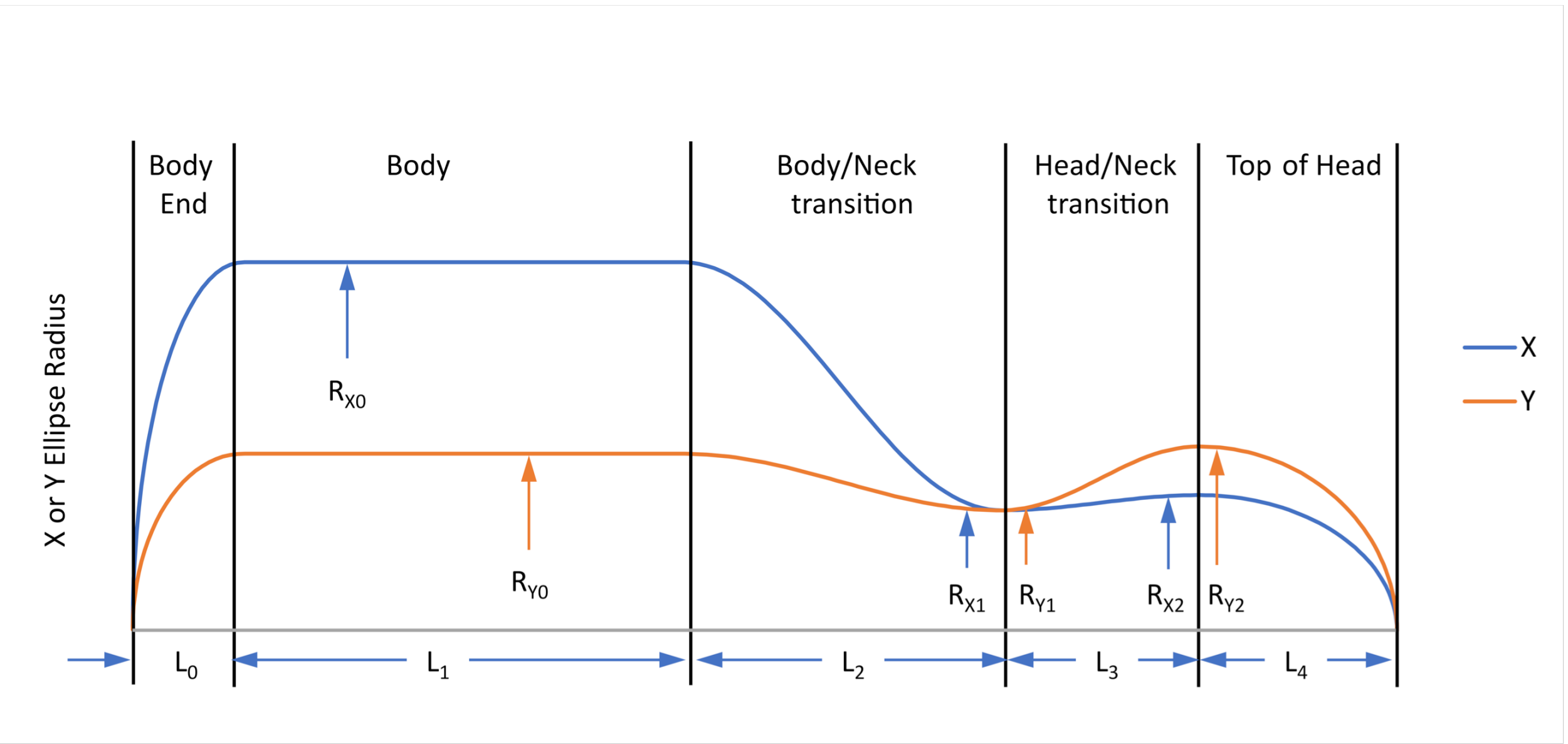

**Figure 1.** Body shape outline in two planes (coronal and sagittal), defined by a set of radii and lengths defining ellipsoidal and elliptic cylindrical regions that represent the body, shoulders and head.

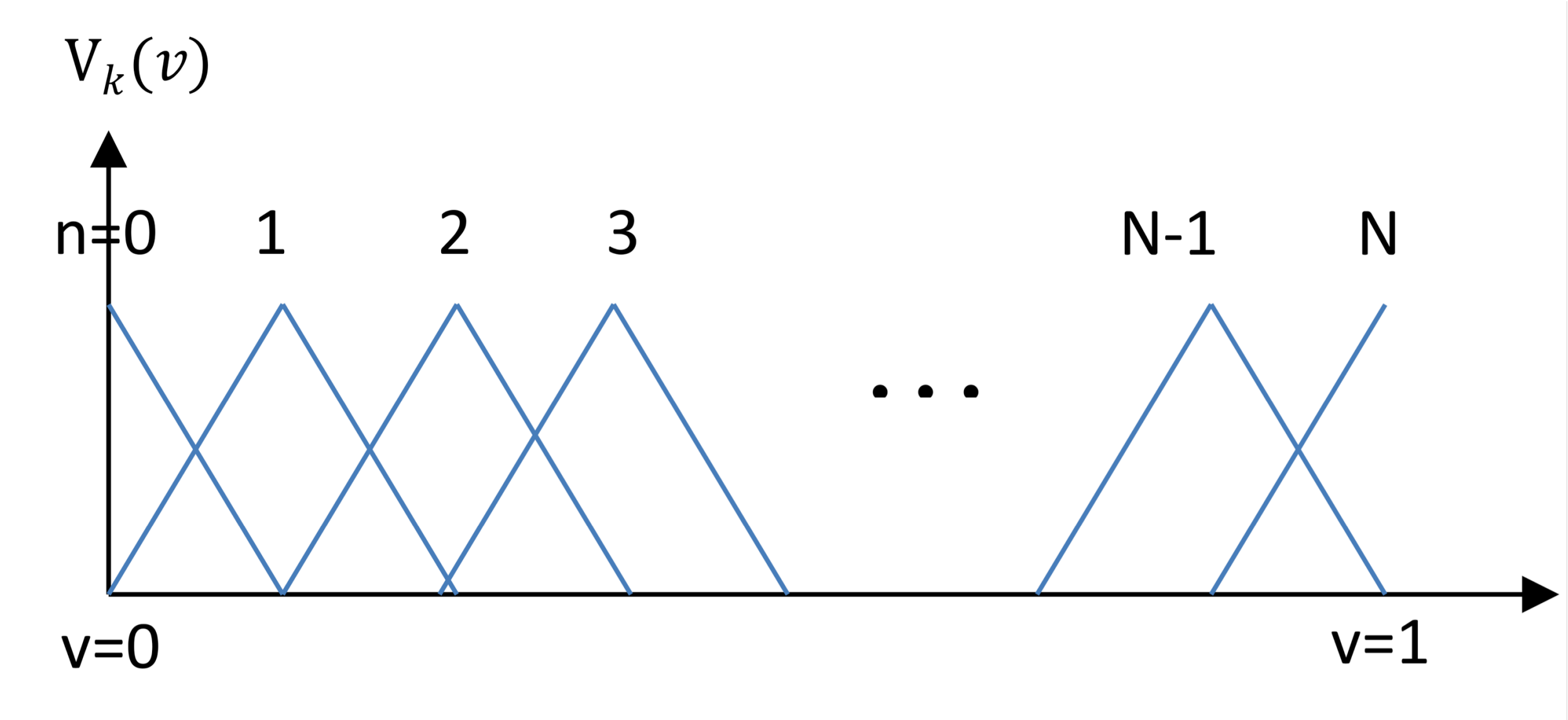

**Figure 2.** Basis function shapes in v direction for calculation of charge distribution.

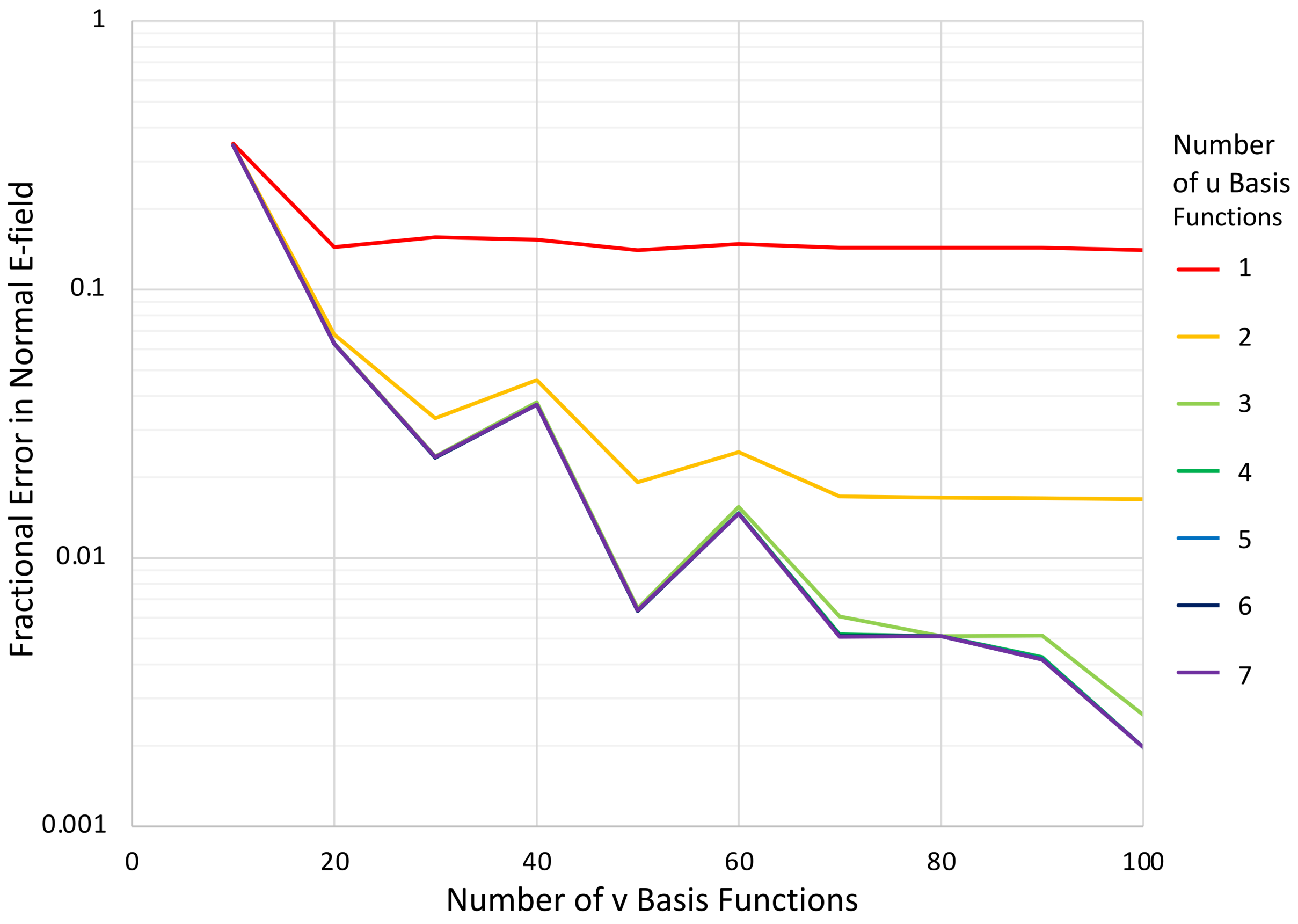

**Figure 3.** Convergence of E-field calculations was assessed by calculating the fractional error in normal E-field, defined as the peak normal component of E-field found anywhere on the body model surface divided by the peak magnitude of E-field found anywhere on the surface. Peak E-field magnitude converges more rapidly than the normal E-field (typical 1 to two order of magnitude faster) due to the fact that the normal E-field adds quadratically to the desired tangential E-field, and hence convergence error is dominated by the error in the normal E-field. Plot shows the convergence behavior of the error in normal E-field for a typical solution as the number of basis functions is increased. Only a few terms in angle (number of u direction basis functions) are required for full convergence. A cylindrically symmetric body model would only require a single term, and the elliptical shape of our body cross section requires only a small number of u direction basis functions for convergence. The v direction needs enough terms to adequately represent the field variation in the z direction; plot shows ~50 terms required to achieve 1% error in the normal E-field.

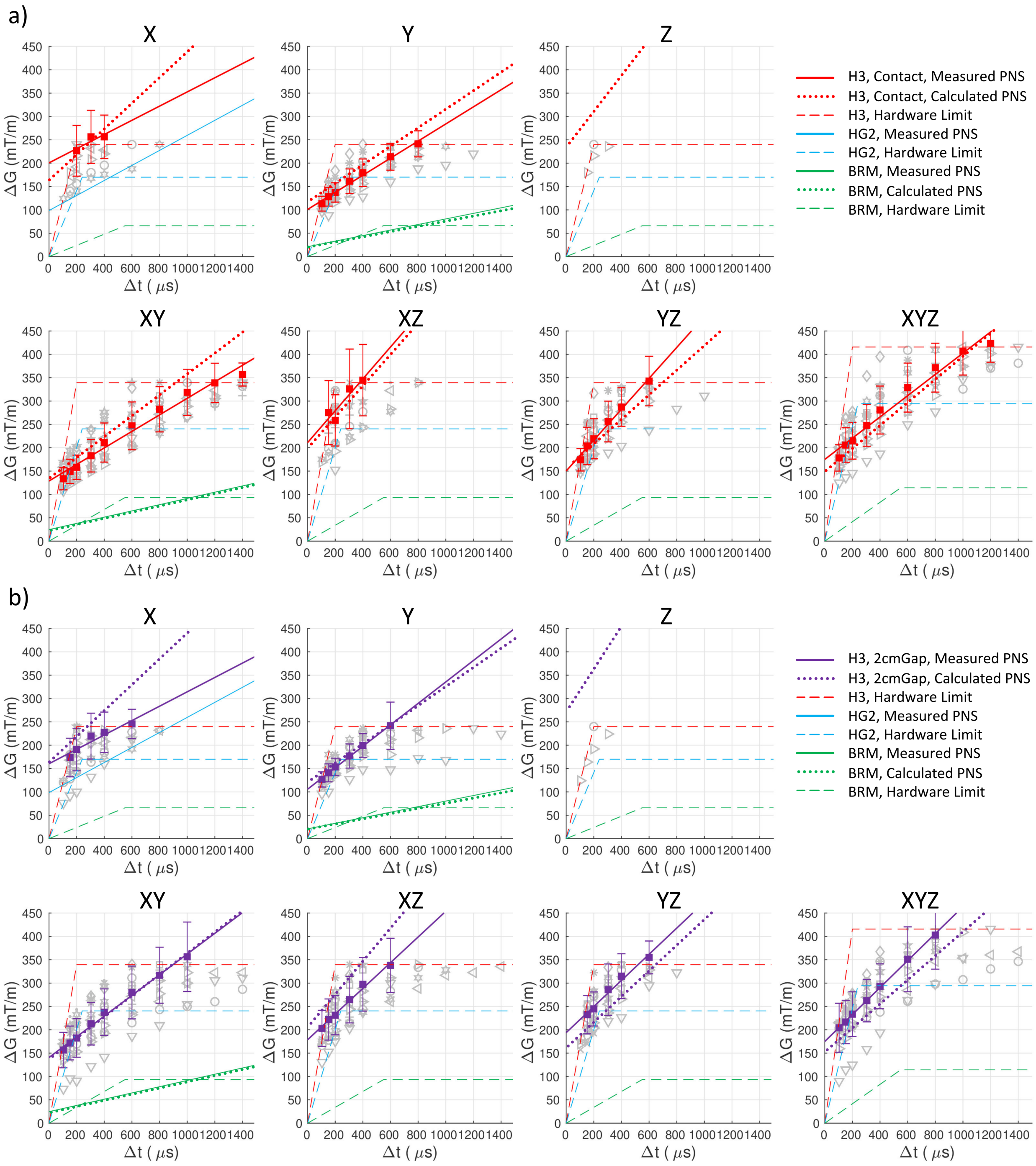

**Figure 4.** Measured and calculated PNS thresholds, superimposed on hardware limits, for seven H3 gradient directions and two body positions. PNS limits are expressed using peak-to-peak standard notation; this means that hardware limits show as twice $G_{max}$ for the particular gradient coil and direction. a) panel corresponds to shoulder-coil contact (brain-centered) position: hardware limits shown as red dashed lines, with individual measured PNS data points (from 15 subjects) shown as open symbols, and logistic regression mean PNS values shown as red points with red linear fits to those mean values. b) panel corresponds to 2cm shoulder-coil gap: individual measured PNS data points shown as open symbols, and logistic regression mean PNS values shown as solid purple points with purple linear fits to those mean values.

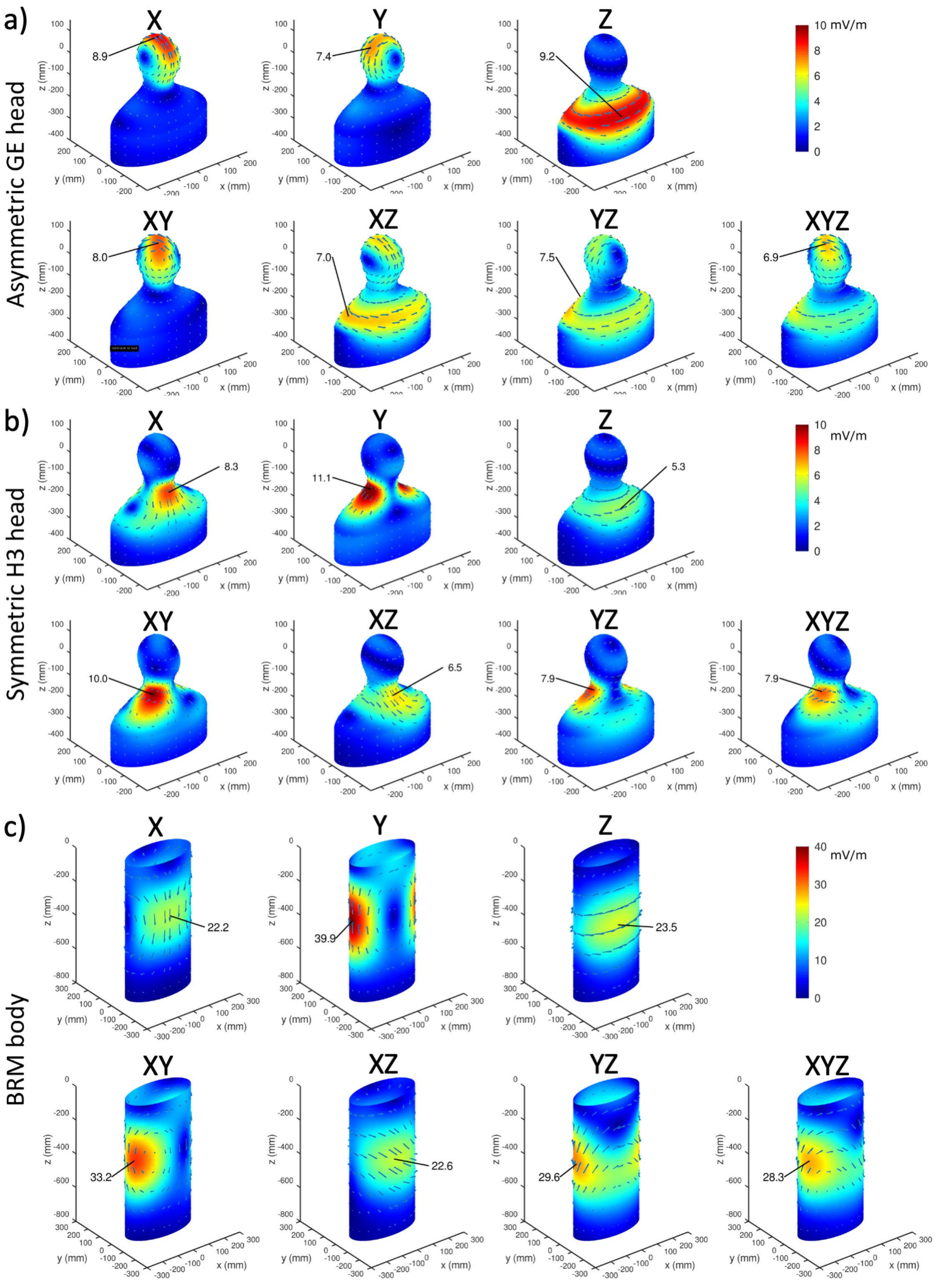

**Figure 5.** Calculated E-field magnitudes in mV/m per T/m/s on surface of patient model for: a) asymmetric GE gradient coil; b) symmetric H3 gradient coil (shoulder contact); and, c) BRM body gradient coil. Numeric value shown in each panel points to the maximum E-field found on the surface. Arrows show E-field direction. For the BRM coil, the body model and fields are symmetric in Z, so only negative Z values are displayed here.